\def\mcr{M_{\rm crust}}
\def\msol{{\rm M}_{\odot}}
\def\nurot{\nu_{\rm rot}}
\def\nurott{\nu_{\rm rot,3}}
\def\nukep{\nu_{\rm K}}
\def\nuisco{\nu_{\rm ISCO}}
\def\req{R_{\rm eq}}

\def\rms{R_{\rm ms}}

\documentclass{aanowy}
\usepackage{graphicx}
\begin{document}
%
%
%
\newcommand{\ddp}[2]{\frac{\partial #1}{\partial #2}}
\newcommand{\ddps}[2]{\frac{\partial^2 #1}{\partial #2 ^2}}

\title{The  crust of  rotating strange quark stars}
\author{J. L. Zdunik \inst{1}
 \and
P. Haensel\inst{1,2}
 \and
E. Gourgoulhon\inst{2}
}
\institute{N. Copernicus Astronomical Center, Polish
           Academy of Sciences, Bartycka 18, PL-00-716 Warszawa, Poland
\and
D\'epartement d'Astrophysique Relativiste et de Cosmologie
-- UMR 8629 du CNRS, Observatoire de Paris, F-92195 Meudon Cedex,
France\\
{\em jlz@camk.edu.pl,  haensel@camk.edu.pl,
  Eric.Gourgoulhon@obspm.fr}}
\offprints{J.L. Zdunik}
\date{Received 12 January 2001/Accepted 9 March 2001}
\abstract{
Calculations of the properties of rotating strange stars
with crusts are performed within the framework of general
relativity. We employ an equation of state (EOS) of strange quark matter based
on the MIT Bag Model with massive strange quarks and lowest order
QCD interactions. The crust is described by the BPS
equation of state.
A significant increase of the stellar radius is found close to
the Keplerian (mass-shedding limit) configuration. This leads to the
disappearance of the
gap between the stellar surface and the
innermost stable circular orbit (ISCO)
 at very high rotation
rates,  for a rather broad range of stellar masses. The Keplerian
configuration for the strange star with crust corresponds to 
values of $J$, $T/W$, $P_{\rm ISCO}=1/\nu_{\rm ISCO}$ which are
 about 20\% smaller than in the
case of bare strange stars. Because the Keplerian configuration is
achieved due to the increase of the stellar oblateness, the
Keplerian frequency (of the rotation) remains almost unaltered.
The lack of the gap close to the Keplerian rotation
could imply a
more stringent limit on $\nu_{\rm ISCO}$,  if the existence of such a
gap is supported by observations, as in the source 4U 1820-30 with
the upper QPO frequency 1.07~kHz. If such a  constraint is taken into
account (mandatory existence of a gap)
the minimum $\nu_{\rm ISCO}$ is about
1~kHz even with the extreme fine tuning
of strange quark matter parameters.
The minimum $\nu_{\rm ISCO}$ is then obtained for the non-rotating
configuration with maximum allowable mass. The maximum frequency in the stable
circular orbit around the strange star with a crust is smaller by about
100~Hz than in the case of a bare strange star.
During the spin-down of a magnetized strange quark star with crust,
the crust matter is absorbed in the equatorial region
by the strange matter core. The deconfinement
of absorbed crust matter is a strongly exothermic process, which
would influence the cooling curve of this compact object.
 \keywords{dense matter -- equation
 of state -- stars: neutron -- stars: rotation
-- stars : binaries}
}

\titlerunning{ The crust of rotating strange stars}
\authorrunning{J.L.~Zdunik et al.}
\maketitle

\section{Introduction}
%
Since the mid-1980, when these exotic
compact stars were introduced
(Witten 1984; first detailed models in Haensel et al. 1986,
Alcock et al. 1986), the existence of strange quark stars
in the Universe remains a
matter of lively debate (for a review, see Madsen 2000).
Recently, these hypothetical objects were invoked in the context of
modeling compact X-ray and gamma-ray sources (Bombaci 1997;
Cheng et al. 1998; Dai \& Lu 1998; Li et al. 1999).
Simultaneously,
several studies, in which exact 2-D general-relativistic
calculations of stationary
rigidly rotating configurations were performed,
  focused on the substantial difference in rapid
rotation of strange quark stars and (normal) neutron stars.
This feature stems from the basic difference  in equations
of state of these two types of compact superdense stars
(Gourgoulhon et al. 1999; Stergioulas et al. 1999; Zdunik
et al. 2000; Gondek-Rosi\'nska et al. 2000).
In particular, the difference in the distribution
of matter in the rapidly rotating models was shown to imply
significant difference in the outer spacetime. Specifically,
qualitative differences in the properties of the innermost stable
circular orbit (ISCO)  around these two
classes of rapidly rotating compact
objects have been found (Stergioulas et al. 1999;
 Zdunik et al. 2000).
The comparison of numerical results for the ISCO (in
particular, the rotation frequency of plasma
blobs on the ISCO, $\nu_{\rm ISCO}$)  with observations
of the kHz quasi-periodic oscillations (QPO) in some of the low-mass
X-ray binaries (LMXB) was used to put constraints on the strange quark
matter (SQM) models.

A strange star  could be covered
 by a   crust of normal matter, a possibility which is
particularly natural in the case of the LMXB (or of a strange star
which went through a LMXB epoch). The problem of
formation and structure of a crust on an accreting
strange star  was studied
by  Haensel \& Zdunik (1991) (see also Miralda-Escud{\'e} et al.
1990). Because of its low mass, typically $\la 10^{-5}~{\rm
M}_\odot$, the effect of the crust on the exterior spacetime is
negligible. However,  due to its thickness,
the crust determines the location of the stellar
surface. As it has been shown in
Zdunik et al. (2000), the presence of the crust could play
an important  role in strange stars rotating close to
Keplerian (mass-shedding limit) frequency. Namely,
the crust undergoes substantial
inflation in the equatorial plane, leading to the
disappearance of the gap between the ISCO and stellar surface.
The crust  mass, thickness and other
macroscopic parameters depend only on the gravitational force exerted
by the SQM core (which is
determined by the  core mass and radius),
and on rotation frequency.

Preliminary, approximate
 studies of the crust on rotating strange stars
 have been  performed
by Glendenning \& Weber (1992), in the framework of Hartle's
(1967) slow rotation  approximation
supplemented by the self-consistency conditions allowing for
determination of the stellar parameters close to the Keplerian
frequencies $\nurot\simeq\nukep$. The results of this approach were
 exploited by Zdunik et al. (2000). They have shown that at some
fixed total baryon mass, $M_{\rm B}$,
 of the strange star, the rotational
increase of the
equatorial radius,
  relative to the nonrotating case,  is
quadratic in $\nu_{\rm rot}/\nu_{\rm K}$. As Zdunik et al. (2000)
have found, the quadratic approximation
described very precisely numerical
results of Glendenning \& Weber (1992),
up to $\nu_{\rm rot}=\nu_{\rm K}$.

In the present paper we present results of  exact (i.e. not based
on the slow rotation approximation) 2-D calculations of
the structure of the crust on rotating strange star.
We  show that the quadratic approximation for
$R_{\rm eq}(\nu_{\rm rot})-R(0)$,
$M_{\rm B,crust}(\nu_{\rm rot})-M_{\rm B,crust}(0)$
is very precise
for  $\nu_{\rm rot}\la 500$ Hz, becomes less
precise with a further increase of $\nu_{\rm rot}$,
and  badly underestimates
the increase of $R_{\rm eq}$ and $M_{\rm B,crust}$
for  $\nu_{\rm rot}\simeq \nu_{\rm K}$.
Consequently, the presence of the solid crust
substantially changes the parameters of the Keplerian configuration,
and the properties of the ISCO for
$\nu_{\rm rot}\simeq \nu_{\rm K}$.

Strange quark stars with a crust could possess a strong magnetic field,
and their magnetosphere may be similar to that of a neutron star.
Therefore, their rapid rotation might be expected to be slowed down
by magnetic braking. Slowing down of rotation would imply a compression
of the crust matter in the equatorial band. As the density at the crust
bottom cannot exceed the critical value $\rho_{\rm b}$ at which
crust matter becomes unstable with respect to irreversible absorption
by the SQM (due to neutron drip or quantum tunneling of nuclei through
the Coulomb barrier), the slowing down of rotation is accompanied by
a decrease in the baryon  mass of the crust $M_{\rm B,crust}$, and
a heating due to energy release accompanying absorption and
deconfinement of baryons (deconfinement heating). We derive
analytical formulae expressing the rate of deconfinement heating
as a function of $\dot\nu_{\rm rot}$ and $\nu_{\rm rot}$.
Deconfinement heating, as well as heating due to non-equilibrium
weak interaction processes in SQM (Cheng \& Dai 1996) will
influence the cooling curve of rotating magnetized
strange quark stars with a crust (Yuan \& Zhang 1999).

In Sect.~2 we briefly present the equation of state SQM and the method
for constructing relativistic 2-D models of stationary rotating
strange quark stars with crust. Numerical results for the parameters
of rapidly rotating strange stars, including the ISCOs, are presented
in Sect.~3. Exact numerical results for the crust on rotating strange
stars are fitted using simple, convenient,
 analytical expressions. The deconfinement heating of spinning-down strange
quark stars with crusts is studied in Sect.~4. Finally, Sect.~5 contains
discussion of our results and  conclusions.
\section{Equation of state and calculation of rotating
strange star models}
\subsection{Equation of state of strange quark matter}
 The differences between the density profiles of a strange star
and a neutron star result from the
basic difference in the EOS of their interiors.
In the case of a rapidly rotating compact object (situation relevant to
 LMXB), the differences between the matter distributions within 
a neutron star and a strange star
may be  expected to  imply differences in  the shape of
rotating objects  and in the spacetime exterior
to the star, and in particular, differences in
the properties of the ISCO.

Our EOS of strange matter, composed of massless u, d quarks, and
massive s quarks, is based on the MIT Bag Model. It involves three
basic parameters: the bag constant, $B$, the mass of the strange
quarks, $m_{\rm s}$, and the QCD coupling constant, $\alpha_{\rm
c}$ (Farhi \& Jaffe 1984, Haensel et al. 1986, Alcock et al.
1986). Our basic  EOS corresponds to standard values of
the Bag Model parameters for strange matter:  $B=56~{\rm
MeV/fm^3}$, $m_{\rm s}=200~{\rm MeV/c^2}$, and $\alpha_{\rm
c}=0.2$ (Farhi \& Jaffe 1984, Haensel et al. 1986, Alcock et al.
1986).  This EOS of strange quark matter will be hereafter
referred to as the SQM1 one. It is the same as that used in 
Zdunik et al. (2000) and Zdunik \& Gourgoulhon (\cite{ZG2001}). 
It yields energy per unit baryon
number at zero pressure $E_0=918.8 ~{\rm MeV} <E(^{56}{\rm
Fe})=930.4~$MeV. For the SQM1 EOS the maximum allowable mass for
static strange stars is $M_{\rm max}^{\rm stat}=1.8~{\rm M_\odot}$.

\subsection{Solid crust of strange quark star} \label{s:solid_crust}
The crust in our model is described by the BPS model of dense matter
below the neutron drip (Baym et al. 1970).
Neutrons are
absorbed by SQM, and therefore the density at the
bottom of the crust, $\rho_{\rm b}$,
 cannot be higher than neutron drip density
$\rho_{\rm ND}\simeq
4\times 10^{11}~{\rm g~cm^{-3}}$. A superstrong outward-directed
electric field separates nuclei of the crust from the
quark-plasma edge, which is necessary for the crust stability
  (Alcock et al. 1986). The gap between nuclei and SQM surface
should be sufficiently large to prevent the absorption
of nuclei by SQM via quantum Coulomb barrier penetration.
If at $\rho_{\rm ND}$ the gap is sufficiently large and the Coulomb
barrier is sufficiently high, then $\rho_{\rm b}=\rho_{\rm ND}$.
Another possibility is that the gap and the Coulomb barrier at
$\rho_{\rm ND}$ can be penetrated by nuclei, and then the stability
conditions are reached at some lower value of $\rho_{\rm b}$. In such a
case, both mass and thickness of the crust are lower than that
corresponding to $\rho_{\rm b}=\rho_{\rm ND}$ (see, e.g., Huang
\& Lu 1997). In what follows, we will assume that
$\rho_{\rm b}=\rho_{\rm ND}$.
The EOS of the crust is joined with that of the SQM core by requiring 
the continuity of pressure at the crust--SQM core interface.
\subsection{Calculation of models of stationary rotating
strange stars with crust} \label{s:code}
%
The general relativistic models of stationary rotating strange stars
have been calculated by means of the code developed by
Gourgoulhon et al. (1999), which relies on
the multi-domain spectral method
introduced by Bonazzola et al. (1998).
With respect to the version of the code used in Gourgoulhon et al. (1999),
Zdunik et al. (2000), Gondek-Rosi\'nska et al. (2000), Zdunik \& Gourgoulhon
(\cite{ZG2001}), we introduce a second domain inside the star to
describe its crust, resulting in a total of four domains. Using the
same notation as in Sect.~3.2 of Gourgoulhon et al. (1999), these
four domains are: D1, which covers the strange quark interior;
D2, which contains the crust; D3, which covers the strong field region
outside the star (and where the ISCO is located) and D4, which extends
 to infinity, thanks to the compactification transformation $u=1/r$.
Following the technique explained in Bonazzola et al. (1998), the
boundary between domains D1 and D2 is forced to coincide with
the transition between the strange quark matter and the crust, so that
the huge density discontinuity between the two regions (cf. Fig.~\ref{dens})
is located at the boundary between the computational domains D1 and
D2 and therefore
does not give rise to spurious oscillations (Gibbs phenomenon).
The number of coefficients used in the spectral expansions (or equivalently
number of grid points) is typically 33 in $r$ and 17 in $\theta$ in each of the
four domains. The global numerical error is evaluated by means of the virial
identities GRV2 (Bonazzola \& Gourgoulhon 1994) and GRV3
(Gourgoulhon \& Bonazzola 1994), this latter being a relativistic
generalization of the classical virial theorem. For instance, for the
configurations depicted in Figs.~\ref{shape} and \ref{dens}, the GRV2
(resp. GRV3) relative error is $6\cdot  10^{-5}$ (resp. $4\cdot 10^{-4}$).
The radius and the frequency of
the marginally stable orbit, $R_{\rm ms}$, and its
frequency $\nu_{\rm ms}$ is determined in a standard way
(see, e.g., Cook et al. 1994, for
the equations to be solved). 

Note that in all the results presented in this article, the equatorial
radius of the star $\req$ is defined as the {\em circumferential radius}
(i.e. the length of the equator, given by the spacetime metric,
divided by $2\pi$). In the nonrotating limit, this equatorial radius
reduces to the surface value of the Schwarzschild coordinate $R$ employed
in the Tolman-Oppenheimer-Volkoff equations.
Note also that all non-equatorial
coordinates (e.g. polar radius, Figs. \ref{shape} and \ref{dens}) are
obtained by rescaling of the radial quasi-isotropic coordinate $r$ 
(see the line element (6) in Gourgoulhon et al. 1999)
by the factor $\req/r_{\rm eq}$.

Let us consider a strange star, rotating
at a frequency $\nu_{\rm rot}$, with equatorial radius $R_{\rm eq}$.
If $R_{\rm ms}>R_{\rm eq}$, then stable orbits exist for $r>R_{\rm ms}$;
the ISCO then has the radius $R_{\rm ms}$ and the
frequency $\nu_{\rm ms}$,
and there is a gap of width $R_{\rm ms}-R_{\rm eq}$ between the
ISCO and the strange star  surface. However, if $R_{\rm ms}<R_{\rm eq}$, then
$R_{\rm ISCO}=R_{\rm eq}$, $\nu_{\rm ISCO}=
\nu_{\rm orb}(R_{\rm eq})$; an accretion disk then extends
down to the strange star
 surface (or, more precisely, joins the stellar surface via a
boundary layer).
 It should be noted that the recent analysis of the
observational data strongly supports the existence of the
marginally stable orbit in the case of 4U 1820-30 (Kaaret et al.
1999).
\section{Results}

\subsection{Global stellar parameters}

In the present section we describe the global parameters of the rotating strange
stars with a crust. We compare these results with those obtained
for bare strange stars and with results of Glendenning \& Weber (1992).

\begin{figure}
\resizebox{\hsize}{!}{\includegraphics[angle=-90]{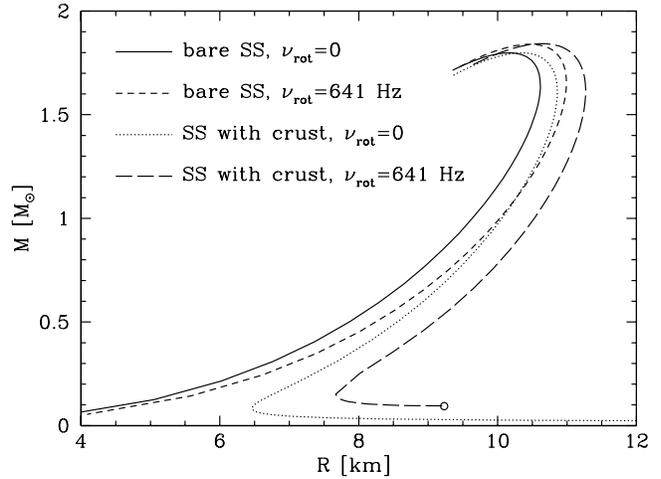}}
\caption{Mass versus (circumferential) equatorial radius for 
nonrotating bare strange stars (SS, solid line), 
 nonrotating SS with crust (dotted), and
SS (bare and with crust) rotating with the maximum observed frequency (641~Hz).
The short-dashed line corresponds to a rotating bare SS and
the long-dashed one to a rotating SS with crust (the 
open circle marks the point at which 641~Hz is the Keplerian frequency).  }
 \label{mrrot}
\end{figure}

In Fig. \ref{mrrot} we present the mass versus radius relation for nonrotating
strange stars and for stars rotating with a period equal to the smallest
measured period of pulsars (1.56~ms, which corresponds to 
a rotation frequency of 641~Hz). 
The last point on the strange star (SS) with crust curve corresponds  to the
Keplerian frequency of this star, equal to 641 Hz. Strange
 stars with smaller mass have thicker crusts and $\nu_{\rm K} <  641$~Hz.

In the next figures we present global parameters of
rotating strange stars (with and without crust) along
some evolutionary sequences
with fixed  total baryon numbers, equal to those of
nonrotating configurations of
given gravitational
masses
 ($M=1.2, 1.4, 1.6, 1.75~\msol$).
Of course, due to the rotation, the actual gravitational
mass of the star increases  with increasing rotation frequency.
It is the baryon mass
$M_{\rm B}$ which is fixed for each sequence
(``evolutionary sequence'').

In these figures, the Keplerian configuration obtained in our
exact calculations  is marked by an open circle.
 For comparison,  we also marked by filled dots
the results of (Zdunik et al. 2000),
obtained using the Glendenning  \& Weber (1992) model for the  crust
on a rotating strange star.
The last point of each curve corresponds to the Keplerian configuration
for  the bare strange star.

\begin{figure}
 \resizebox{\hsize}{!}{\includegraphics[angle=-90]{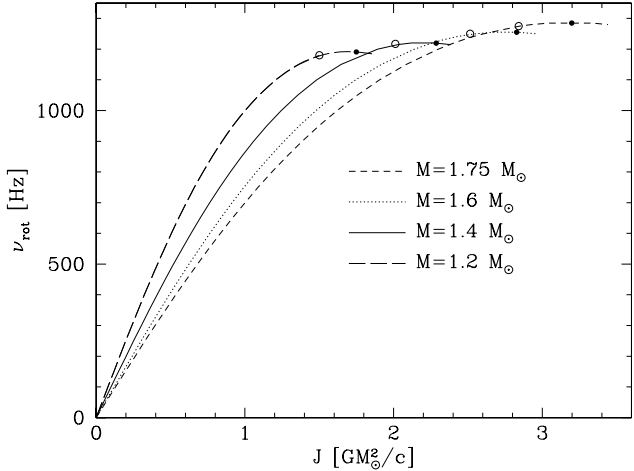}}
\caption{Rotation frequency versus  angular
momentum of SS along constant baryon number sequences. 
Each sequence is labeled by the gravitational mass $M$ of its nonrotating
configuration. 
The last points of these curves correspond to
the Keplerian configurations for bare SS: {\em filled dots:} Glendenning-Weber
approximation of the rotating crust; {\em open dots:} exact calculations
of the Keplerian frequency.}
 \label{fj}
\end{figure}

In Fig. \ref{fj} we present $\nurot$ as a function of the angular momentum
of the star $J$.
We see that for a bare strange star  the Keplerian configuration is
not achieved via  the
increase of $\nurot$. On the contrary,
close to this configuration there exists
 a region for
which $\nurot$ is a slightly decreasing function of $J$
: this corresponds to spin up by the angular  momentum loss.
 Therefore, $\nu^{\rm max}_{\rm rot}>\nu_{\rm rot}(J_{\rm max})\equiv
\nu_{\rm K}$.

This effect turns out to  be suppressed by the existence of the crust.
Although the crust does not change in practice the curve $\nurot(J)$,
the radius of a rotating star with a crust is significantly
larger, setting the maximum value of $J$ to be smaller by $\sim 20\%$
than for a bare strange star. However,
due to the flatness of the function  $\nurot(J)$ for
high $J$, the Keplerian frequency is nearly the same as in the case of bare SS.

\begin{figure}
 \resizebox{\hsize}{!}{\includegraphics[]{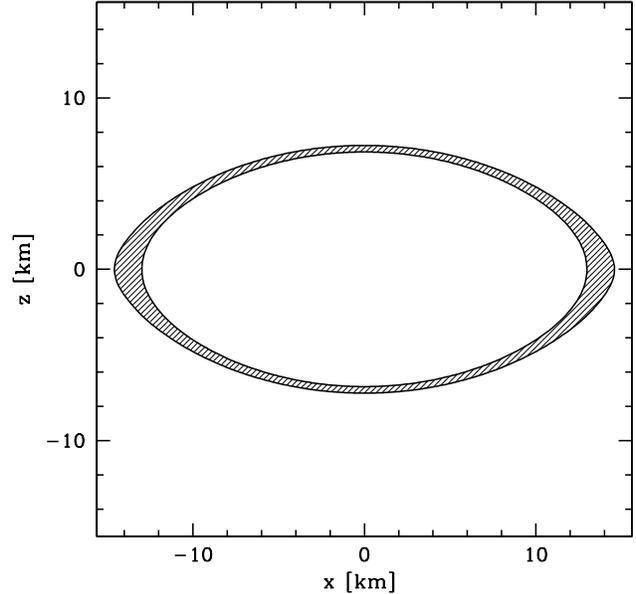}}
\caption{The edge of the SQM core and the crust surface for a
strange quark star with crust (baryon mass $M_{\rm B}=1.63~
{\rm M}_\odot$), rotating at $\nu_{\rm rot}=1210$~Hz.
The Keplerian frequency $\nu_{\rm K}=1217~$Hz (see Sect.~\ref{s:code} 
for the definition of the coordinates used in this plot). 
}
 \label{shape}
\end{figure}

\begin{figure}
 \resizebox{\hsize}{!}{\includegraphics[angle=-90]{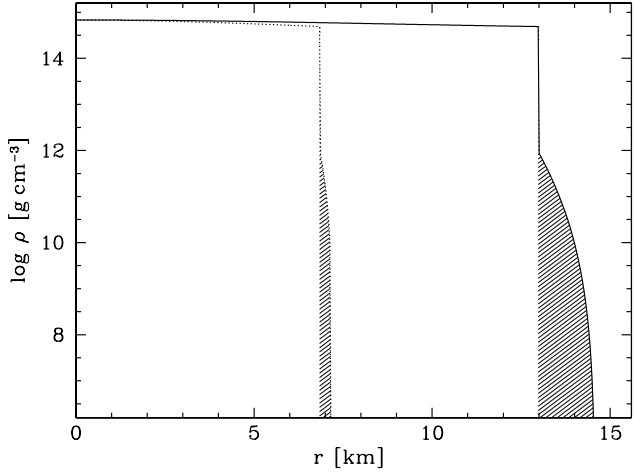}}
\caption{Logarithm of density versus the radial coordinate 
(defined in Sect.~\ref{s:code}) along the polar (dotted line) and equatorial
(solid line) directions. The strange star model is the same as 
in Fig.~\ref{shape}. The shaded areas correspond to the crust.}
 \label{dens}
\end{figure}

\begin{figure}
 \resizebox{\hsize}{!}{\includegraphics[angle=-90]{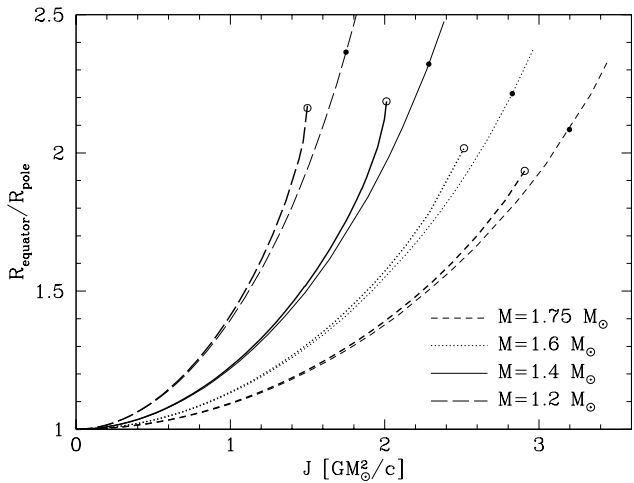}}
\caption{Oblateness of the star presented as the ratio of the equatorial
to polar radius, versus stellar angular momentum $J$,
along some constant baryon number sequences. 
Each sequence is labeled by the gravitational mass $M$ of its non-rotating
configuration. 
Thin lines correspond to bare
strange stars.
}
 \label{oblj}
\end{figure}

The reason why $\nurot(J)$ does not increase close to
$J_{\rm max}$ is explained in Fig.
\ref{oblj}. The increase of $J$ results in flattening of the
star ($\req$ increases, so $J$ increases without change in $\nurot$).
In other words,
the Keplerian configuration is achieved due to the increase in
the equatorial radius related to the deformation of
the star.
We see that the strange star with a crust is more oblate than
bare SS for the same $J$. This is due to the fact that
the crust is the most deformed part of the star.
This effect is visualized in Figs. \ref{shape} and \ref{dens},
in which we show the effect of rotation at
$\nu_{\rm rot}=1210~$kHz on the structure of a strange quark star
of $M_{\rm B}=1.63~{\rm M}_\odot$. As apparent in Fig.~\ref{shape},
the crust thickness at the equator is much greater than at the pole
(about five times). Correspondingly, the density gradient in the polar
ring of the crust is dramatically less steep than at the pole
(see Fig.~\ref{dens}). Generally, the increase  of  $R_{\rm eq}/R_{\rm pole}$
with the total stellar angular momentum $J$ depends on the baryon
mass of the star: the lower the value of $M_{\rm B}$, the steeper the
increase of $R_{\rm eq}/R_{\rm pole}$ with increasing $J$.
This is visualized in  Fig. \ref{tcrf}.

\begin{figure}
 \resizebox{\hsize}{!}{\includegraphics[angle=-90]{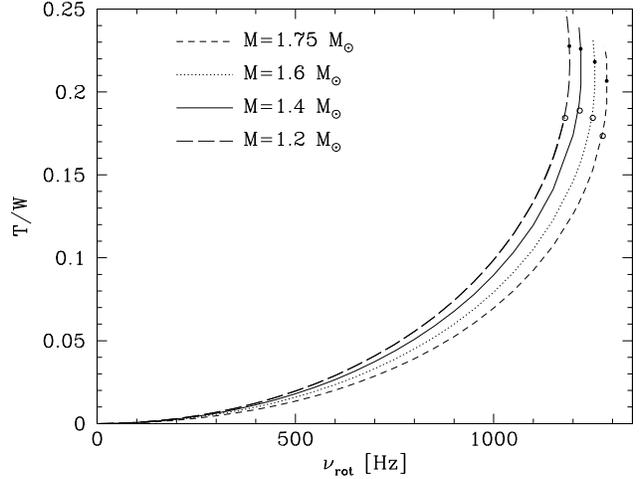}}
\caption{$T/W$ ratio as a function of $\nurot$ along some constant
baryon number sequences. 
Notations as in Fig \ref{oblj}.}
 \label{twf}
\end{figure}

The global parameter which is relevant from the point of view of
the stability of rapidly rotating  stars with respect to
spontaneous triaxial deformations
is the ratio of kinetic to gravitational energy
$T/W$ (Shapiro \& Zane \cite{Shapza}).
In Fig. \ref{twf} we have plotted $T/W$ versus rotation frequency,
for four evolutionary sequences of rotating strange stars with some fixed
baryon number.
The differences in the maximum value of $T/W$
between the cases of bare strange stars and strange stars
 with crust are quite
significant.
For strange stars with crust the maximum value
 $T/W \approx 0.18\div 0.19$, which is more than
 20\% smaller than in the case of
bare strange stars, with maximum $T/W \approx 0.25$.

\subsection{Innermost stable circular orbit}

\begin{figure}
 \resizebox{\hsize}{!}{\includegraphics[angle=-90]{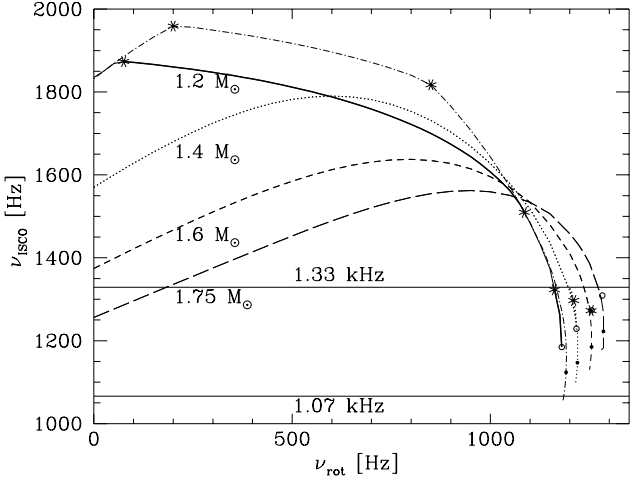}}
\caption{Orbital frequency of the ISCO versus the rotation frequency of strange
stars,  for the SQM1 EOS of strange matter.
Each curve corresponds to a fixed
baryon mass, equal to that of a static strange star of
gravitational mass indicated by a label.
Along each curve, the angular momentum increases from $J=0$ (static
configuration) to $J_{\rm max}$ (Keplerian limit). Open circles
 correspond to Keplerian configurations of strange stars with crust
(filled circles -- approximation of the crust based on the
results of Glendenning and Weber (1992)).
Segments below the filled circles can be reached
only by bare strange stars. Asterisks are
defined by the condition $R_{\rm ms}=R_{\rm eq}$:
below  these asterisks in the high frequency region
and between asterisks for 1.2 $\msol$ case at intermediate rotation
rate (thick part of the curve) $R_{\rm ms}<R_{\rm eq}$
and the ISCO is determined by the stellar radius $R_{\rm eq}$.
Dash-dotted line corresponds to the bare strange star
of $1.2\,\msol$. The difference
with respect to the star with crust in the region $\sim 100-1100$~kHz
is due to the fact that for bare strange stars
the marginally stable orbit disappears at higher $\nurot$
and reappears at lower $\nurot$ leading to the $\nuisco$
 larger than for SS with crust by $\sim100$~Hz.}
\label{isco}
\end{figure}

Previous calculations suggested that
the gap between the ISCO and the stellar surface
exists for Keplerian configurations  within a rather
broad range of stellar masses (Stergioulas et al. 1999,
Zdunik et al. 2000). Our calculations, presented in Fig.~\ref{isco},
indicate that for strange stars with crusts the gap exists only
very close to the maximum baryon mass. For lower masses (even $1.6 \,\msol$
for the SQM1 model for which $M^{\rm stat}_{\rm max}=1.8\,\msol$)
there is no gap close
to the Keplerian configuration .
Thus if the observational data support the existence of the gap,
the theoretical lower limit on $\nuisco$ for rapidly
rotating strange stars will be higher (asterisks in Fig. \ref{isco}).

For bare strange stars  the gap exists up to the Keplerian frequency.
Even in the case of smaller mass stars for which $\rms<\req$
for intermediate frequencies (e.g. $M=1.2~\msol$) the gap
reappears at high rotation rates.
In the extreme case of low mass bare strange stars (for which
relativistic effects are negligible) the gap appears close to
Keplerian frequency  due to the large oblateness of the
star (Zdunik \& Gourgoulhon \cite{ZG2001}).
As a result, the frequency of the ISCO at the Keplerian limit is
much smaller than for the nonrotating star
(especially for low-mass stars).

Of course, the smaller the mass,
the larger the difference between bare strange stars and strange
stars with a crust,
the size of the crust being a decreasing function of $M$.
For low mass strange stars with crusts ($M<1.1\,\msol$)
the gap between the surface and the marginally stable orbit
vanishes independent of the rotation rate.
In the intermediate mass range ($\sim 1.1\div 1.3 \,\msol$)
the gap exists for a nonrotating star, then disappears for
larger rotation rates and reappears
at high frequencies.
In this case the difference
between a bare strange star and a star with a crust is clearly
visible (Fig. \ref{isco}).
Due to the existence of the crust, the region of $\nurot$
in which the ISCO is determined by the stellar surface (i.e. $\rms<\req$)
is larger and thus the frequency $\nuisco$ is significantly
smaller because the orbit of ISCO
corresponds then to the equatorial radius and not to the mass
of the star (as in the case of the marginally stable orbit).
As a result, the maximum possible value of the frequency of the particle
in a stable circular orbit around the strange
star with crust  (which is determined by the point $\rms=\req$)
is smaller by about 100 Hz than for bare SS.

For relatively massive strange stars
 with a crust ($\sim 1.3\div 1.6\,\msol$),
the gap exists for all rotation rates except those
very close to $\nu_{\rm K}$. The condition
$\rms=R_{\rm eq}$
defines then the minimum ISCO frequency consistent with the constraint
of the existence of the marginally stable orbit
($\sim 1.3$~kHz for $1.2<M/{\rm M}_\odot<1.6$ in the case of the SQM1 EOS,
asterisks in Fig. \ref{isco}).

Only for stars with a mass close to the maximum for nonrotating
configurations ($\sim 1.6\div 1.8\,\msol$) is the marginally stable orbit
located above the stellar surface at Keplerian rotation rate.

As the Keplerian rotation corresponds to the minimum of
$\nu_{\rm ISCO}$, the existence of the crust increases these values
by $100\div 200$~Hz compared to bare strange stars. As one can expect,
 $\nu_{\rm ISCO}$ is then equal to $\nukep$ since ISCO is determined by
the equatorial radius of the star rotating with Keplerian frequency.

However it should be stressed that the lower limit on $\nuisco$ consistent
with the assumption that a marginally stable orbit exists (i.e.
$\rms>R_{\rm eq}$) is significantly higher and is equal to
$\sim 1.27$~kHz for SQM1 model and $1.1$~kHz for fine tuning
of strange matter parameters, leading
to the SQM2 EOS (Zdunik et al. 2000).
For this last EOS
$\nu_{\rm ISCO}<1$~kHz can be reached by bare SS for very rapid rotation.
In this case
the condition $R_{\rm ms}>R_{\rm eq }$ eliminates rapidly rotating models
 with $\nu_{\rm ISCO}< 1.1$~kHz, and the
condition  $\nu_{\rm ISCO}=$1.07 kHz with a non-zero gap
can be fulfilled only by
slowly rotating strange stars  very close to
maximum mass.
Contrary to bare strange stars, when even for low mass stars rotating
at the Keplerian limit, the frequency of the orbiting particle is limited by
the marginally stable orbit and is approximately constant for
a wide range of stellar masses (Zdunik \& Gourgoulhon \cite{ZG2001}),
for strange stars with a
crust the marginally stable orbit does not exist at the Keplerian limit
and the innermost stable orbit coincides with the radius of the star
(except for the mass close to $M_{\rm max}$),
which rapidly increases as the mass decreases, leading to the decrease of
the frequency on ISCO.

\subsection{Parameters of the crust}
Two parameters of the crust, relevant for our considerations, are
its baryon mass $M_{\rm B,crust}$ and 
its equatorial thickness $t_{\rm eq}$, defined in terms of the
circumferential radius introduced in Sect.~\ref{s:code}.
In the static case these quantities will be
denoted as $M^0_{\rm B,crust}$ and $t_0$, respectively. In the static case
we will additionally consider crust contribution to the stellar
gravitational mass $M^0_{\rm crust}$. 
This quantity does not have a clear invariant meaning, reflecting the
fact that gravitational energy cannot be localized in general 
relativity. Accordingly, $M^0_{\rm crust}$ should be considered
only as the difference between the total gravitational mass $M$ and the value
at the interior/crust boundary of the metric 
coefficient $m(R)$ (``enclosed mass'') which enters the
Oppenheimer-Volkoff equations. 

Let us start with the simplest case of a static strange
star. For $M>{\rm M}_\odot$, the ratio
$t_0/R$ is  small, and this enables us to derive
approximate expressions resulting from the Oppenheimer-Volkoff
equations of hydrostatic equilibrium
\begin{eqnarray}
\mcr^0&=&{4\pi R^4 P_{\rm b}\over GM}
\left(1-{2GM\over Rc^2}\right)~,\nonumber\\
t_0&=&{\gamma\over \gamma-1} {R^2c^2\over GM}
{P_{\rm b}\over \rho_{\rm b}
c^2}~,
\label{M.t0.stat}
\end{eqnarray}
where $\rho_{\rm b}$ and $P_{\rm b}$ are the
density and pressure at the bottom of the
crust, and $\gamma$ is the (average) adiabatic index of the outer
crust.

\begin{figure}
 \resizebox{\hsize}{!}{\includegraphics{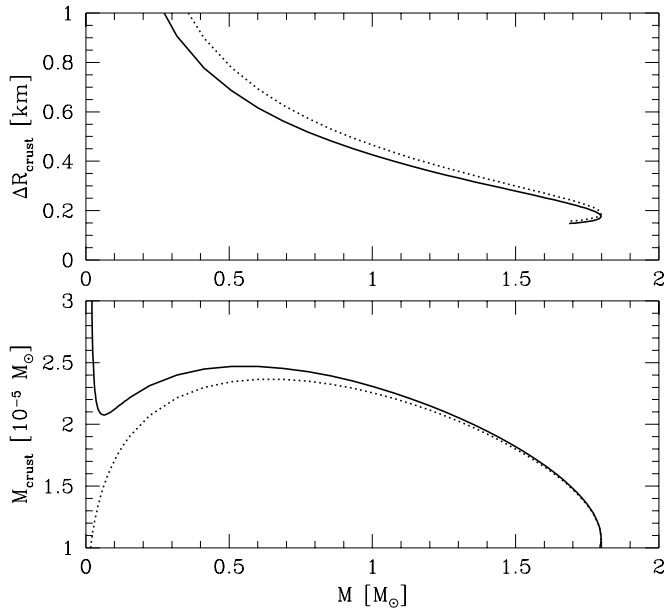}}
\caption{Accuracy of the approximations (\ref{tmapp}) for the thickness
and mass of the crust of a nonrotating strange star. 
{\em Solid line:} exact solution of the
Oppenheimer-Volkoff equation; {\em dotted line:} approximation 
given by formula~(\ref{tmapp}).}
\label{crustap}
\end{figure}

For the BPS  model (Baym et al. 1970) and $\rho_{\rm b}$ equal to the neutron
drip density, we have
$P_{\rm b}=7.8\,10^{29}~{\rm dyn~cm^{-2}}$,
${P_{\rm b}/ \rho_{\rm b} c^2}=0.002$, $\gamma=1.28$.
This yields
%
\begin{eqnarray}
t_0&=&0.65{R_{6}^2\over M_*} \left(1-0.295
{M_*\over R_{6}}\right)~{\rm km}~,\nonumber\\
\mcr^0&=&3.7\,10^{-5}{R_{6}^4\over M_*}\left(1-0.295
{M_*\over R_{6}}\right)~\msol~,
\label{tmapp}
\end{eqnarray}
where $R_{6}$ is the stellar radius in units $10^6$~cm and
$M_*=M/\msol$.
The above formulae reproduce the exact results with an error lower
than $6\%$ in the case of $t_0$ and $4\%$ for $\mcr$ for strange
stars with mass between $1~\msol$ and $2~\msol$ (maximum mass).
A comparison between approximate and exact values  is
displayed  in Fig. \ref{crustap}.

\begin{figure}
 \resizebox{\hsize}{!}{\includegraphics{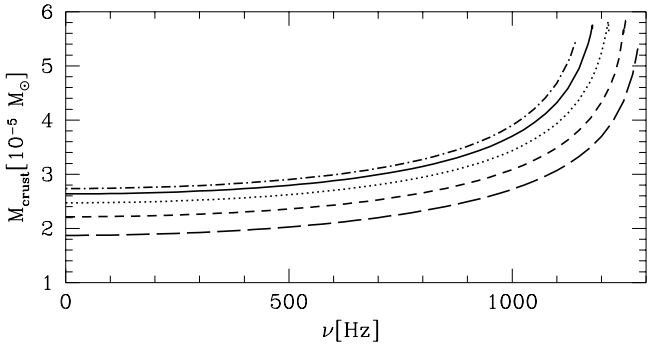}}
\caption{Baryon mass of the crust
 as a function of rotational frequency, along sequences 
of constant baryon number. The gravitational masses $M$
of the nonrotating configuration of each sequence are 
 1.75, 1.6, 1.4,.1.2, 1$M_\odot$ (from bottom
to top). Note that less massive stars have a larger crust 
(see also Fig \ref{crustap}). }
\label{mbcrf}
\end{figure}

\begin{figure}
 \resizebox{\hsize}{!}{\includegraphics{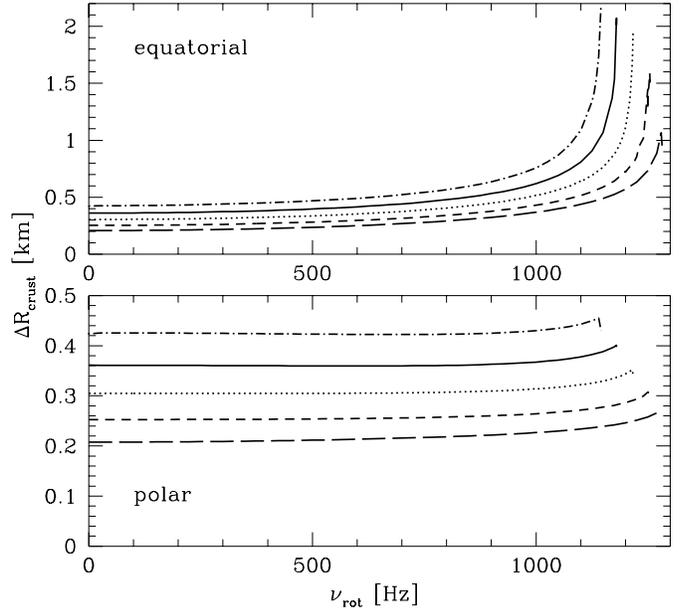}}
\caption{Same as Fig.~\ref{mbcrf} but for the thickness of the crust 
(equatorial and polar). Note that the two plots are shown using different scales
and that for a nonrotating star equatorial and polar thicknesses
are equal.}
\label{tcrf}
\end{figure}

Rotation will increase both the mass and the equatorial
thickness of the crust.  Results of  Glendenning and Weber (1992)
for  $t_{\rm eq}$ as a function of  $\nu_{\rm
rot}$ (at fixed value of $M_{\rm B}$)
could be very precisely  reproduced by a
simple formula $t_{\rm eq}(\nu_{\rm
rot})=t_0\cdot [1+0.7(\nu_{\rm rot}/\nu_{\rm K})^2]$, where
$\nu_{\rm K}$ is the Keplerian (mass shedding) frequency of the strange star
 (Zdunik et al. 2000). A similar formula reproduces very well
the rotational increase of the baryon mass of the crust.

Exact results for the crust on the static and rotating strange stars
 are presented in Figs.~\ref{mbcrf} and \ref{tcrf}.
The mass of the crust which can be supported
by the rotating strange star
can even be 2.5 times larger than in the case of a nonrotating
configuration. Another very important effect is a strong
oblateness of the star,
resulting in large $\req$.
As a result, the crust at the equator is much thicker than at the
pole (see Figs.~\ref{shape} and \ref{dens}).

The mass and thickness of the crust can be approximated by a quadratic
function of rotation frequency
up to about 500 Hz.
Thus the results of Glendenning and Weber (1992) describe very well
the parameters of the crust of the star rotating with intermediate
frequencies (say below one half of the Keplerian one).
For higher values  of $\nu_{\rm rot}$, both
crust mass and equatorial thickness
increase rapidly and
we have to use a polynomial including terms of much
higher (sixth to eighth)  order in $\nu_{\rm rot}$.
The approximate formulae are found to be
\begin{eqnarray}
M_{\rm B,crust}&=&M^0_{\rm B,crust} (1+0.24\,\nurott^2+0.16\,\nurott^8)\,,
\label{Mcrust.rot} \\
t_{\rm eq}(\nu_{\rm rot})&=&t_0\cdot (1+0.4\,\nurott^2+0.3\,\nurott^6)\,,
\label{Rcrust.rot}
\end{eqnarray}
where $\nurott=\nurot/10^3$ Hz.
Equation (\ref{Mcrust.rot}) works quite well nearly
up to the Keplerian frequency, but the equatorial radius
increases very rapidly as one approaches the Keplerian configuration.
In view of this, Eq. (\ref{Rcrust.rot}) can be safely used below 1~kHz
(i.e. about 80\% of the Keplerian value), where the change of the
equatorial thickness is $\sim 100\%$. The difference between
our quadratic approximate formula
and the fitting results of
 Glendenning \& Weber (1992) is a few percent.

Another interesting effect of rotation on
the crust structure deserves comment.
As Glendenning \& Weber (1992)  pointed out,
the polar thickness also increases (albeit weakly)
due to the decrease in the gravitational
force (resulting from the oblateness of mass distribution)
at the pole. This effect is confirmed by our calculations
(see bottom panel of Fig. \ref{tcrf})

\section{Deconfinement heating  in spinning-down strange
quark stars with crust}
The spin-down of a magnetized strange star implies a decrease
of its oblateness. This process is particularly pronounced
in the strange star crust. This change of shape  is
 opposed by the elastic
stress build-up in the Coulomb lattice of nuclei.
The crust is relatively ``soft'' with respect
to shear deformations, because its shear modulus constitutes
only a few percent of the compression modulus (Haensel 1997, Haensel 2001).
Therefore, in a zeroth approximation, the shape of the shrinking
crust of a spinning-down strange star can be modeled by the equilibrium
shape of a fluid. The spin-down implies some compression of the bottom
matter elements of the crust, most pronounced in the equatorial
belt. Consequently, as soon as the density exceeds the critical
value $\rho_{\rm b}$, the bottom crust elements are engulfed
by the SQM core. We note that if $\rho_{\rm b}=\rho_{\rm ND}$
then the absorption of crust matter is triggered by the neutron
drip out of nuclei, while for $\rho_{\rm b}<\rho_{\rm ND}$
the absorption proceeds via quantum tunneling of nuclei through
the Coulomb barrier, separating them from the SQM core
(see Sect.~\ref{s:solid_crust}).

A typical spin-down rate for a  relatively young
$1000~$yr old pulsar (like Crab) is
${\dot P}\sim 10^{-13}~$s/s, where $P$ is the pulsar period. Using
the relation
$\nu_{\rm rot}{\dot\nu}_{\rm rot}=-{\dot P}/P^3$, and
expression  (\ref{Mcrust.rot}) for the
  instantaneous baryon mass
of the crust, we calculate the rate at which
the baryon mass of the crust decreases  due to the strange star
spin-down:
%
\begin{eqnarray}
{\dot M}_{\rm B,crust}&\simeq& -9.6~10^{14}\;
{ M^0_{\rm B,crust}\over 10^{-5}~{\rm M}_\odot}\;
\left({P\over 10~{\rm ms}}\right)^{-3}\cr
&~&\times {\vert{\dot P}\vert\over 10^{-13}~{\rm s/s}}
~~{\rm g\, s}^{-1}~.
\label{Mcrust.dot}
\end{eqnarray}
After crossing the SQM surface, nucleons (those in nuclei and
dripped neutrons) dissolve into $u$ and $d$ quarks (via strong interactions),
which will be followed by  the strangeness-changing equilibration process
$u+d\longrightarrow  s +u$ via weak interaction. The whole process
is expected to be strongly exothermic, with an energy release per
unit absorbed baryon number $q_{\rm dec}\simeq 10 - 40~$ MeV, its
specific value depending  on the assumed SQM model (Farhi and Jaffe 1984).
Using Eq.~(\ref{Mcrust.dot}), we   estimate
the total heating rate at the crust-SQM interface as
%
\begin{eqnarray}
{Q}_{\rm dec}&\simeq& 9~10^{33}
\cdot \left({P\over {\rm 10~ms}}\right)^{-3}
\cdot {\vert{\dot P}\vert \over {\rm 10^{-13}}{\rm s/s}}\cr
&\times&{M^0_{\rm B,crust}\over 10^{-5}~{\rm M}_\odot}\cdot
{q_{\rm dec}\over {\rm 10~MeV}}~{\rm erg\, s}^{-1}~.
\label{Q.dec}
\end{eqnarray}
%
The typical deconfinement heating, estimated above, may therefore
exceed the neutrino losses (luminosity) from the SQM core of a
1000 years old (or older)  strange star with a crust,
$L_\nu\sim 10^{31}~ (R/10~{\rm km})^3\;
(T/10^7~{\rm K})^6~{\rm erg/s}$, and therefore could
significantly influence (namely  - delay) the cooling of this compact object
(Yuan \& Zhang 1999).
We mention that the deconfinement heating, described above, is
much more efficient than the roto-chemical heating, resulting from
deviations of the composition of the SQM core of a spinning-down strange star
from beta equilibrium  and studied in  (Cheng \& Dai 1996).
\section{Summary  and conclusion}
We calculated exact 2-D models of stationary rotating strange stars
of masses $1.2-1.75~{\rm M}_\odot$, covered
with a crust of normal matter, up to the Keplerian limit.
Despite its low mass  ($\sim 10^{-5}~{\rm M}_\odot$) and small
thickness (a few hundred meters), the solid crust plays an important
role for these hypothetical objects. We considered sequences
of uniformly rotating strange stars of a fixed baryon mass,
and increasing values of angular momentum $J$,   and
we found that due to the large deformation near the Keplerian limit,
the decrease of the maximum  frequency, compared to that for
bare strange stars, is very small. Close to the Keplerian limit,
the stellar moment of inertia increases  rapidly with rotation
frequency, and this results in a weak dependence  of
the maximum frequency of uniform rotation,  $\nu_{\rm max}$, on the
 value of $J_{\rm max}$, which in the presence of the crust
is  lower by as much as twenty percent compared to that  of bare
strange stars.
On the contrary,  the
 effect of the presence of the crust on the
properties of the innermost stable circular stable orbit is important,
and consists of about fifteen percent increase in $\nu_{\rm ISCO}$
at the Keplerian limit, as well as the disappearance of the gap between
the ISCO and stellar surface for a broad range of stellar masses
and rotation frequencies.  The solid crust also lowers by as much as
twenty percent the maximum
  $T/W$ ratio
which can be reached by a rotating strange star, from  $0.23-0.25$ to
$0.18-0.20$.

 Both the mass of the crust and
its equatorial thickness increase with the rotation frequency. We have
shown that this increase
is very well approximated by simple and precise
analytical formulae up to 80\% of the Keplerian frequency.
For a rotation
frequency below one third of the maximum, the increase with respect
to the static case  is quadratic in the rotation frequency. However,
close to the Keplerian limit, the dependence is much more rapid, which is
consistent with the characteristic behavior of $J(\nu_{\rm rot})$,
mentioned in the previous paragraph.

\begin{acknowledgements}
This research
was partially supported by the KBN grant No. 5P03D.020.20.
The numerical calculations have been performed on computers purchased
thanks to a special grant from the SPM and SDU departments of
CNRS. The main results reported in the present paper were obtained during  
a visit of one of the authors (J.L.Z) at DARC, Observatoire de Paris-
Meudon, within the framework of the  CNRS/PAN
program  Jumelage Astrophysique.
\end{acknowledgements}


\begin{thebibliography}{} 

\bibitem[1986]{Alcock86}
Alcock C., Farhi C.E., Olinto A., 1986,
ApJ,  310, 261


\bibitem[1971]{BPS}
Baym G., Pethick C., Sutherland P., 1971, ApJ, 170, 299

\bibitem[1997]{Bombaci97}
Bombaci I., 1997, Phys. Rev. C,  55, 1587

\bibitem[1994]{bona94}
Bonazzola S., Gourgoulhon E., 1994,
Class. Quantum Grav., 11, 1775


\bibitem[1998]{Bona98}
Bonazzola S., Gourgoulhon E., Marck J.A., 1998,
Phys. Rev. D,  58, 104020

\bibitem[1999]{Bulik99}
Bulik T., Gondek-Rosi{\'n}ska D., Klu{\'z}niak W. 1999,
A\&A,  344, L71

\bibitem[1996]{Cheng96}
Cheng K.S.,  Dai Z.G., 1996,
ApJ, 468, 819

\bibitem[1998]{Cheng98}
Cheng K.S.,  Dai Z.G., Wei D.M.,  Lu T., 1998,
 Science,  280, 407


\bibitem[1994]{CST94b}
Cook G.B., Shapiro S.L., Teukolsky S.A., 1994, ApJ, 424, 823

\bibitem[1998]{Dai98}
Dai Z.G., Lu T., 1998, Phys. Rev. Lett., 81, 4301



\bibitem[1984]{FJ84}
Farhi E., Jaffe R.L., 1984, Phys. Rev. D,  30, 2379

\bibitem[1992]{GW92}
Glendenning N.K., Weber F., 1992, ApJ,  400, 647


\bibitem[2000]{Gon00}
Gondek-Rosi{\'n}ska D., Bulik T., Zdunik L., Gourgoulhon E., 
Ray S., Dey J., Dey M., 2000, A\&A, 363, 1005

\bibitem[1994]{Gour94}
Gourgoulhon E., Bonazzola S., 1994, Class. Quantum Grav., 11, 443

\bibitem[1999]{Gour99}
Gourgoulhon E., Haensel P., Livine R., Paluch E., Bonazzola S., Marck
J.A., 1999, A\&A , 349, 851

\bibitem[1997]{H97}
Haensel P., 1997, In: Marck J.-A., Lasota J.-P. (eds.)
Relativistic Gravitation and Gravitational Radiation. 
Cambridge University Press, p. 129

\bibitem[2001]{H2001}
Haensel P., 2001, In: Blaschke D., Glendenning N.K., Sedrakian A. (eds.)
Physics of Neutron Star Interiors. Springer Verlag
(in press)

\bibitem[1986]{HZS86}
Haensel P., Zdunik J.L., Schaeffer R., 1986,
A\&A, 160, 121

\bibitem[1991]{HZ91}
Haensel P., Zdunik J.L., 1991,
Nucl. Phys. B (Proc. Suppl.),  24B, 139

\bibitem{Hartle67} Hartle J.B., 1967, ApJ, 150, 1005

\bibitem[1997]{HuangL97}
Huang Y.F., Lu T., 1997, A\&A, 325, 189

\bibitem[1997]{Kaaret97}
Kaaret P., Ford E.C., Chen K., 1997,
ApJ Lett., 480, L27

\bibitem[1999]{Kaaret99}
Kaaret P., Piraino S., Bloser P.F., Ford E.C., Grindlay J.E.,
Santangelo A., Smale A.P., Zhang W., 1999,
ApJ Lett., 520, L37




\bibitem[1999]{Li99}
Li X.-D., Bombaci I., Dey M., Dey J., van den Heuvel E.P.J., 1999,
Phys. Rev. Lett. 83, 3776


\bibitem[1999]{Madsen99}
Madsen J., 1999, In: Cleymans~J. (ed.)
Hadrons in Dense Matter and Hadrosynthesis. Lecture
Notes in Physics, Springer-Verlag, p.\ 162



\bibitem[1990]{Miralda90}
Miralda-Escud{\'e} J., Haensel P., Paczy{\'n}ski B., 1990,
ApJ, 362, 572


\bibitem[1998]{Shapza}
Shapiro S.L, Zane S., 1998, ApJS, 117, 531

\bibitem[1998]{Shibata99}
Shibata M., Sasaki M., 1998, Phys. Rev D, 58, 104011

\bibitem[1999]{Sterg99}
Stergioulas N., Klu{\'z}niak W., Bulik T., 1999,
A\&A, 352, L116




\bibitem[1984]{Witt84} Witten E., 1984, Phys. Rev. D,  30, 272

\bibitem[1999]{Yuan1999}
Yuan Y.F., Zhang J.L., 1999, A\&A, 344,  371

\bibitem[2001]{ZG2001} Zdunik J.L., Gourgoulhon E., 2001,
Phys. Rev. D, 63, 087501



\bibitem{ZHGG00}
Zdunik J.L., Haensel P., Gondek-Rosi\'nska, Gourgoulhon E., 2000,
A\&A, 356,  612

\end{thebibliography}
\end{document}